\documentclass[a4paper,12pt]{article}
\usepackage{cite}
\usepackage[ref]{overcite}
\usepackage{graphicx}
\begin{document}
\baselineskip=0.8truecm
\title{\bf Ray splitting in paraxial optical cavities}

\author{G. Puentes, A. Aiello, J. P. Woerdman \\
 {\em Huygens Laboratory, Leiden University, P.O. Box 9504,
Leiden, The Netherlands}}
\maketitle

\begin{abstract}
We present a numerical investigation of the ray dynamics in a
paraxial optical cavity when a ray splitting mechanism is present.
The cavity is a conventional two-mirror stable resonator and the
ray splitting is achieved by inserting an optical beam splitter
perpendicular to the cavity axis. We show that depending on the
position of the beam splitter the optical resonator can become
unstable and the ray dynamics displays a positive Lyapunov
exponent.
\end{abstract}

\begin{flushleft}
 \end{flushleft}
\thispagestyle{empty}

\section{INTRODUCTION}
A beam splitter (BS) is an ubiquitous optical device in wave
optics experiments, used e.g., for optical interference,
homodyning, etc. In the context of geometrical optics, light rays
are split into a transmitted and reflected ray by a BS. Ray
splitting provides an useful mechanism to generate chaotic
dynamics in pseudointegrable \cite{Kohler97a} and soft-chaotic
\cite{Couchman92a,Hentschel02a} closed systems. In this paper we
exploit the ray splitting properties of a BS in order to build an
open paraxial cavity which shows irregular ray dynamics as opposed
to the regular dynamics displayed by a paraxial cavity when the BS
is absent.

\begin{figure} \begin{center}
\includegraphics[angle=0,width=6.5truecm]{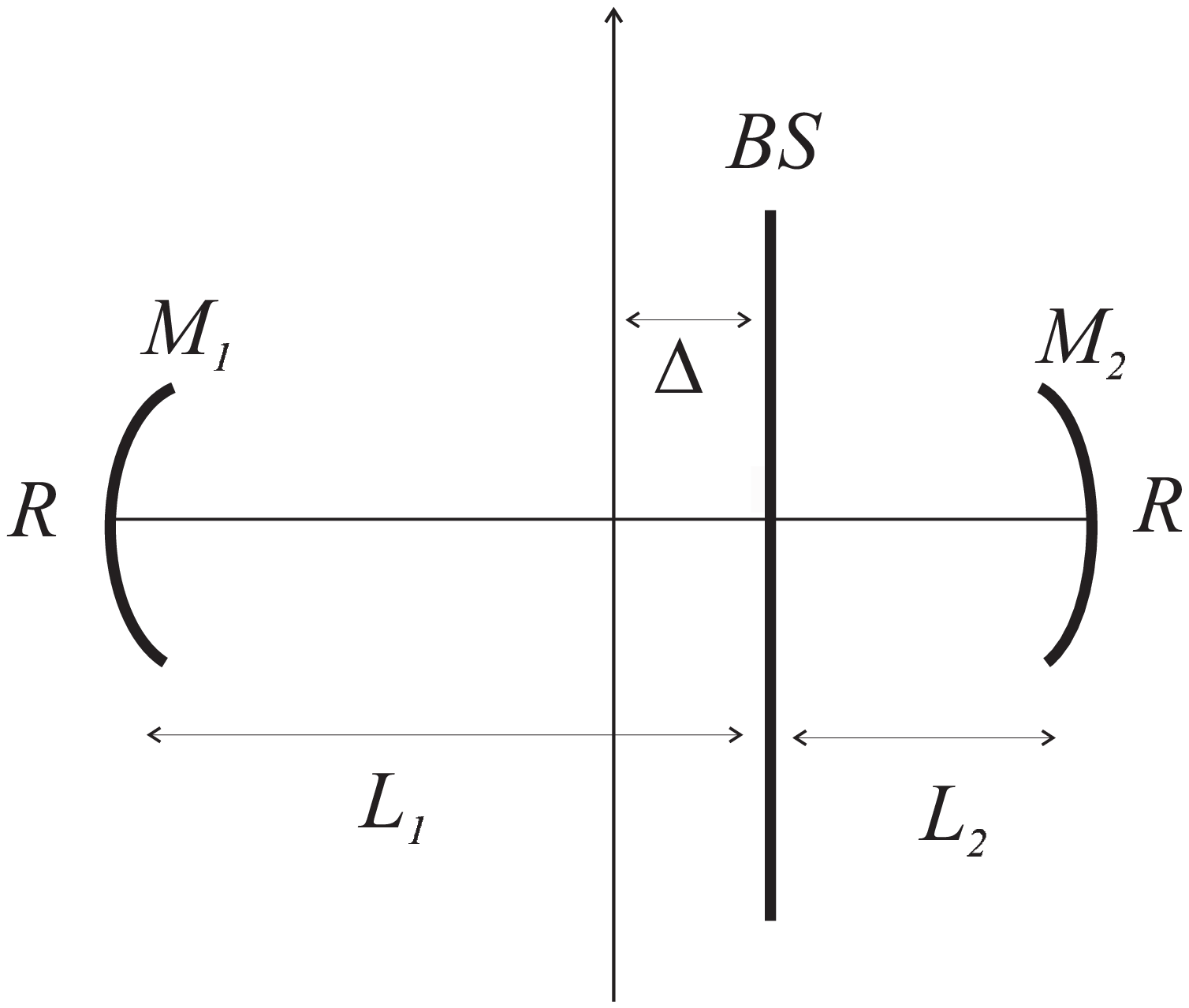}
\caption{\label{fig:1} Schematic diagram of the cavity model. Two
subcavities of length $L_{1}$ and $L_{2}$ are coupled by a BS. The
total cavity is globally stable for $L=L_1+L_2 < 2 R$. $\Delta=L_1
- L/2$ represents the displacement of the BS with respect to the
center of the cavity.}
\end{center} \end{figure}

\begin{figure} \begin{center}
\includegraphics[angle=0,width=6.5truecm]{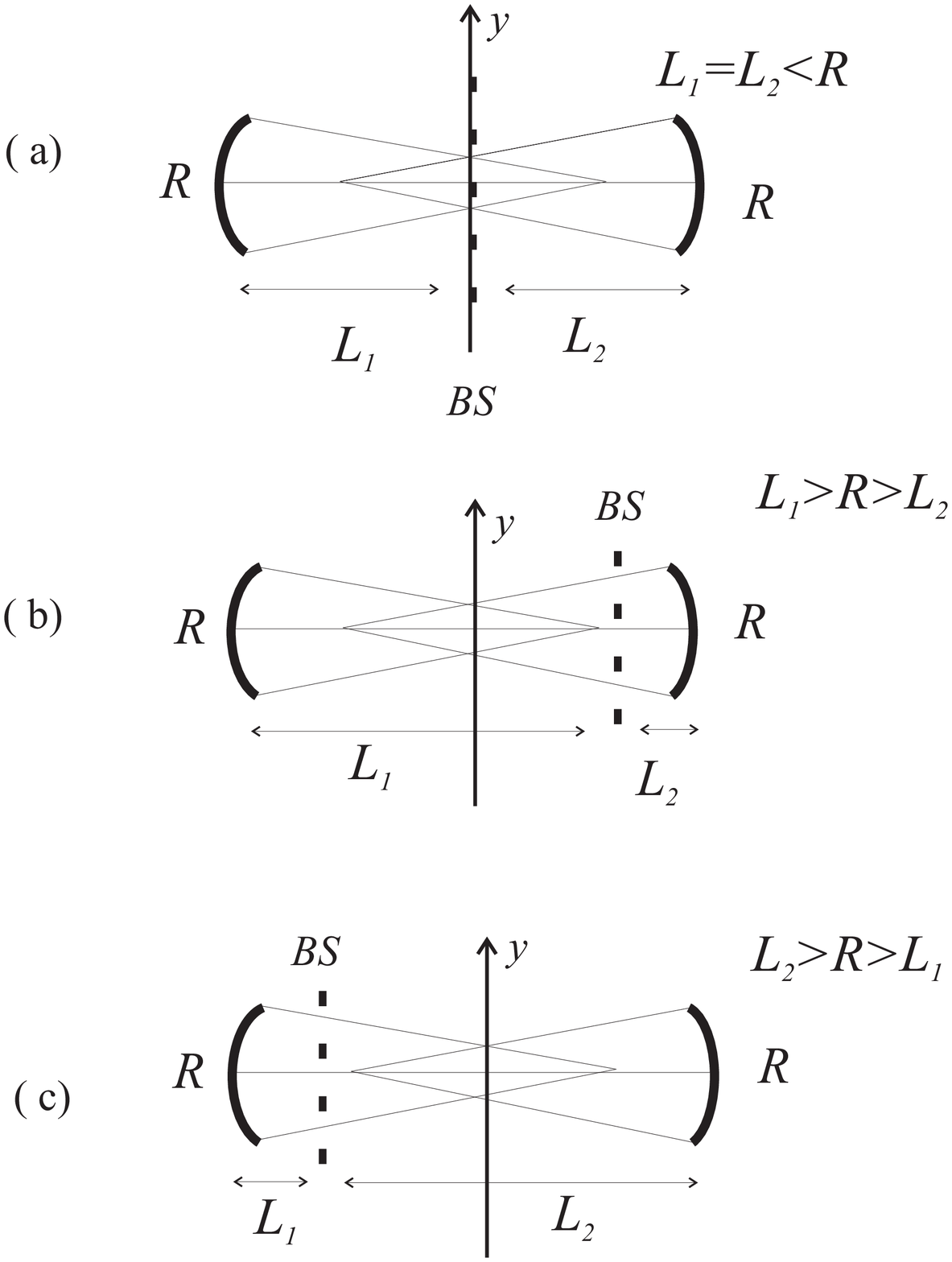}
\caption{\label{fig:3} The different positions of the beam
splitter determine the nature of the subcavities. In (a) the BS is
in the middle, so the 2 subcavities are stable, in (b) the left
cavity is unstable and the right one is stable, and (c) the
unstable (stable) cavity is on the left (right)  (b).}
\end{center} \end{figure}

Optical cavities can be classified as \emph{stable} or
\emph{unstable} depending on the focussing properties of the
elements that compose it \cite{SiegmanBook}. An optical cavity
formed by 2 concave mirrors of radii $R$ separated by a distance
$L$ is stable when $L<2R$ and unstable otherwise. If a light ray
is injected inside the cavity through one of the mirrors it will
remain confined indefinitely  inside the cavity  when the
configuration is stable but it will escape after a finite number
of bounces  when the cavity is unstable (this number depends on
the degree of instability of the  system). Both stable and
unstable cavities have been extensively investigated since they
form the basis of laser physics \cite{SiegmanBook}. Our interest
is in a composite cavity which has both aspects of stability and
instability. The cavity is  made by two identical concave mirrors
of radii $R$ separated by a distance $L$, where $L<2R$ so that the
cavity is globally stable. We then introduce a beam splitter (BS)
inside the cavity, oriented perpendicular to the optical axis (Fig
\ref{fig:1}). In this way the BS defines two subcavities. The main
idea is that depending on the position of the BS the left (right)
subcavity becomes \emph{unstable} for the reflected rays when
$L_1$ ($L_2$) is bigger than $R$, whereas the cavity as a whole
remains always \emph{stable} ($L_1+L_2<2R$) (Fig. \ref{fig:3}).

\begin{figure} \begin{center}
\includegraphics[angle=0,width=6.5truecm]{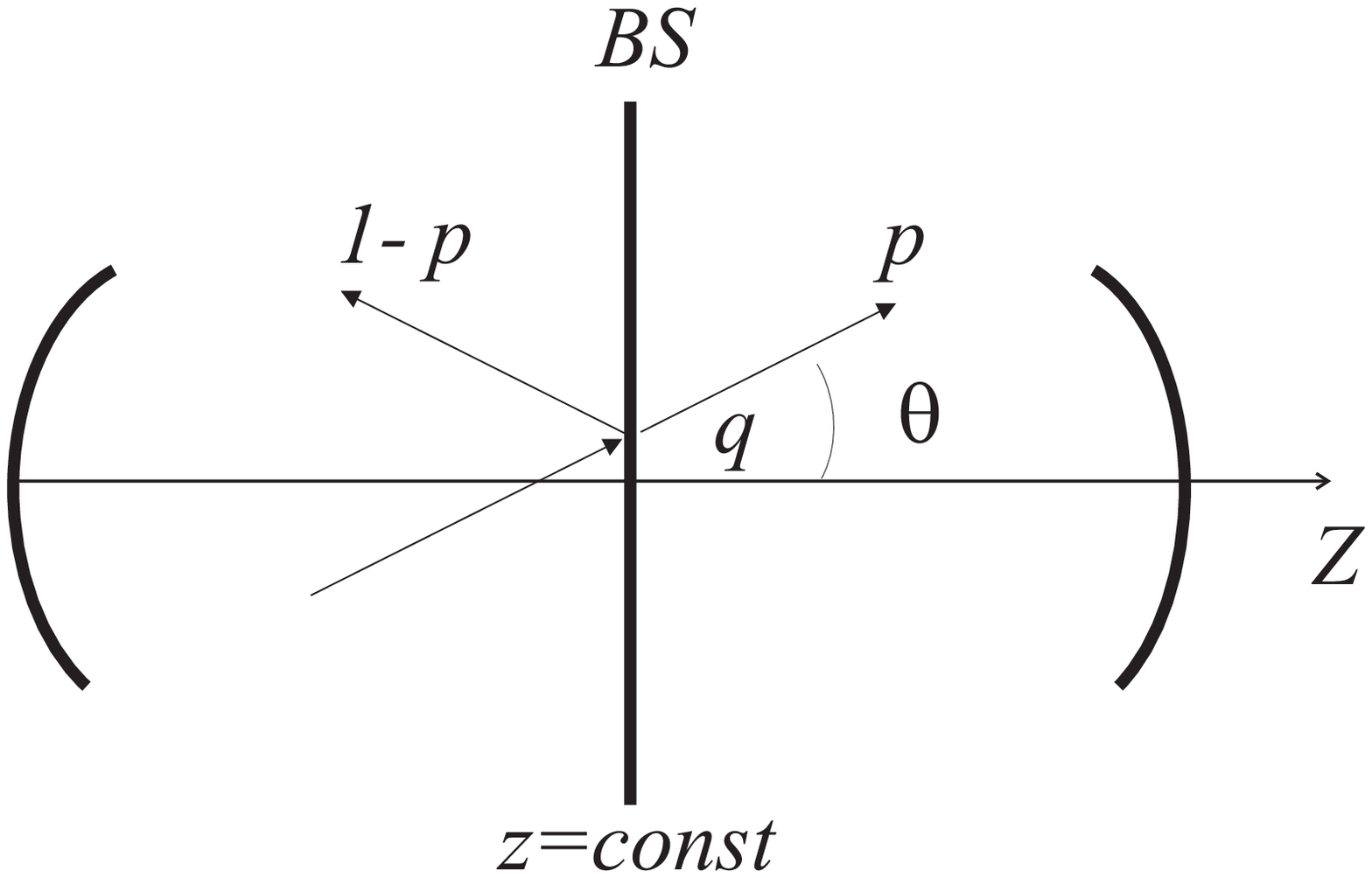}
\caption{\label{fig:7} A ray on a reference plane ($z=const$)
perpendicular to the optical axis ($Z$) is specified by two
parameters: the height $q$ above the optical axis and the angle
$\theta$ between the direction of propagation and the same axis.
When a ray hits the surface of the BS, which we choose to coincide
with  the reference plane, it can be either reflected or
transmitted with equal probability. For a $50\%/50\%$ beam
splitter $p=1/2$.}
\end{center} \end{figure}

Our motivation to address this system originates in  the
nontrivial question whether there will be a balance between
trapped rays and escaping rays. The trapped rays are those which
bounce infinitely long in the stable part of the cavity, while the
escaping ones are those which stay for a finite time, due to the
presence of the unstable subcavity. If such balance exists it
could eventually lead to transient chaos since it is known in
literature  that instability (positive Lyapunov exponents) and
mixing (confinement inside the system) form the
skeleton of chaos \cite{Cvitanovic02}.\\
The BS is modelled as a stochastic ray splitting element
\cite{Couchman92a} by assuming the reflection and transmission
coefficients as random variables. Within the context of wave
optics this model corresponds to the neglect of all interference
phenomena inside the cavity; this would occur, for instance when
one injects inside the cavity a wave packet (or cw broad band
light) whose longitudinal coherence length  is very much shorter
than the smallest characteristic length of the cavity. The
stochasticity  is implemented by using a Monte Carlo method to
determine whether the ray is transmitted or reflected by the BS
\cite{Couchman92a}. When a ray is incident on the ray splitting
surface of the BS, it is either transmitted through it with
probability $p$ or reflected with probability $1-p$, where we will
assume $p=1/2$, i.e. we considered  a $50\%/50\%$ beam splitter
(Fig \ref{fig:7}). We then follow a ray and at each reflection we
use a random number generator with a uniform distribution to
randomly decide whether
to reflect or transmit the incident ray.\\
Our system bears a close connection with the stability of a
periodic guide of paraxial lenses as studied by  Longhi
\cite{Longhi}. While in his case  a \emph{continuous} stochastic
variable $\epsilon_{n}$ represents a perturbation  of the periodic
sequence along which rays are propagated, in our case we have a
\emph{discrete} stochastic parameter $p_{n}$ which  represents the
response of the BS  to an incident ray. As will be shown in
section II, this stochastic parameter can take only two values,
either 1 for transmitted rays or -1 for reflected ray; in this
sense, our  system (as displayed in Fig.\ref{fig:2}) is a
surprisingly simple realization of  a bimodal stochastic paraxial
lens guide.

The structure of the paper is as follows. In section II we
describe the ray limit, and the paraxial map or ABCD matrix
associated with rays that propagate very close to the axis of the
cavity. In section III we present the results of the numerical
simulations  for the paraxial map associated with our ray optical
system; these simulations are based on standard numerical tools
developed in non-linear dynamics theory. Finally, in section IV,
we detail the conclusions of our work.

\section{Ray Dynamics and the Paraxial Map}

The time evolution of a laser beam inside a cavity can be
approximated classically by using the ray optics limit, where the
wave nature of light is neglected. Generally, in this limit the
propagation of light in a uniform medium is described by rays
which travel in straight lines, and which are either sharply
reflected or refracted when they hit a medium with a different
refractive index. To fully characterize the trajectory of a ray in
a strip resonator or in a resonator with rotational symmetry
around the optical axis, we choose a reference plane $z=constant$
(perpendicular to the optical axis $\hat{z}$), so that a ray is
specified by two parameters: the height $q$ above the optical axis
and the angle $\theta$ between the trajectory and the same axis.
Therefore we can associate a ray of light with a two dimensional
vector $\vec{r}=(q,\theta)$. This is illustrated in the two mirror
cavity show in Fig. \ref{fig:7}, where the reference plane has
been chosen to coincide with the beam splitter (BS). Given  such a
reference plane $z$, which is  also called Poincar\'{e} Surface of
Section  (SOS) \cite{OttBook},  a round trip (evolution between
two successive reference planes) of the ray inside the cavity can
be calculated by the monodromy matrix $M_{n}$, in other words
$\vec{r}_{n+1}=M_{n}\vec{r}_{n}$, where the index $n$ determines
the number of round trips. The monodromy matrix $M_{n}$ describes
the linearized evolution of a ray that deviates from a reference
periodic orbit. A periodic orbit is said to be \emph{stable} if
$|$Tr$M_{n}| < 2$. In this case nearby rays oscillate back and
forth around the stable periodic orbit with bounded displacements
both in $q$ and $\theta$. On the other hand when $|$Tr$M_{n}|\geq
2$ the orbit is said to be unstable and rays that are initially
near this reference orbit become more and more displaced from it.\\

\begin{figure} \begin{center}
\includegraphics[angle=0,width=7.5truecm]{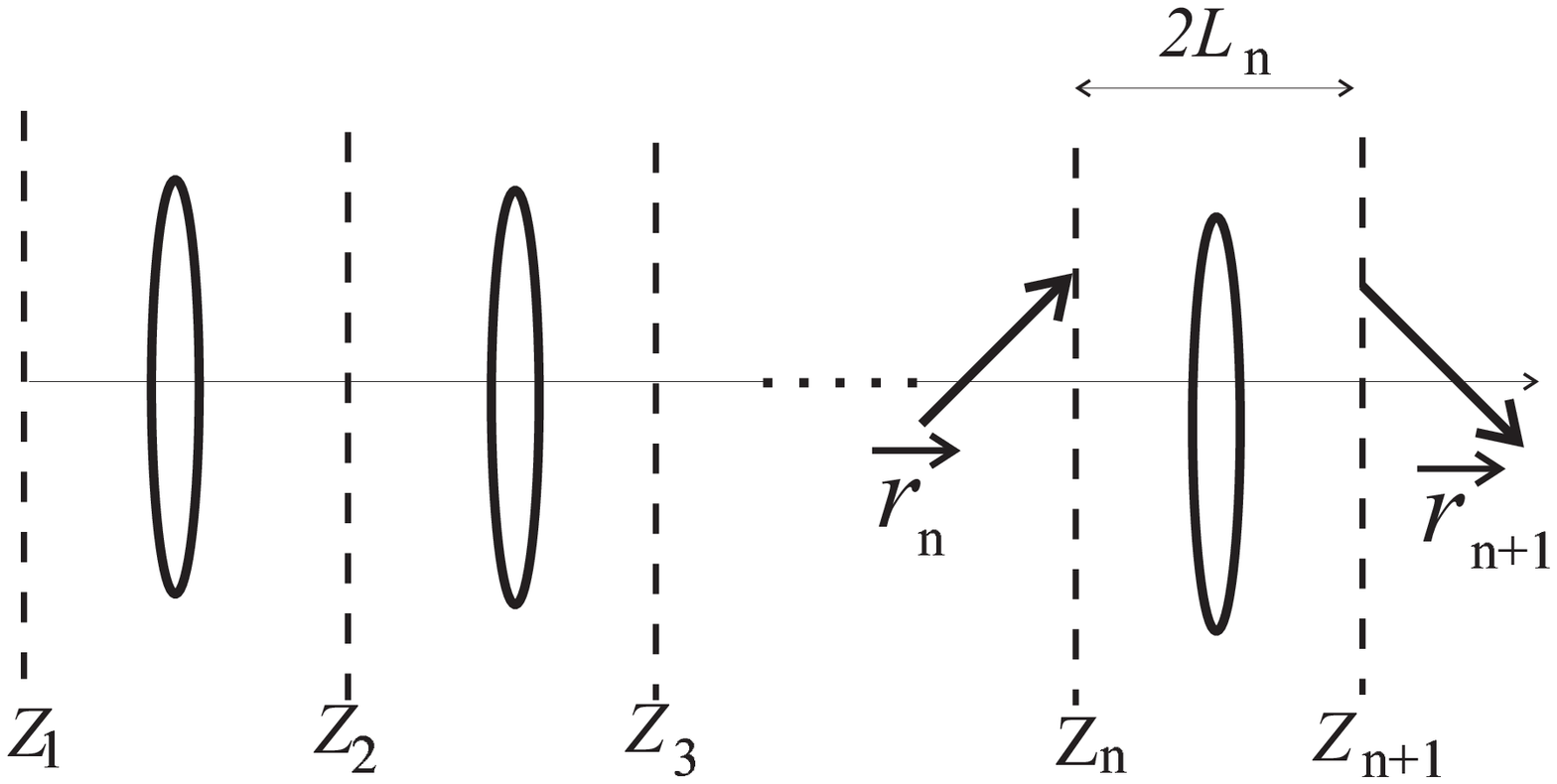}
\caption{\label{fig:2} A ray bouncing inside an optical cavity can
be represented by a sequence of lenses of focus $f=2/R$, followed
by a free propagation over a distances $L_{n}$. Due to the
presence of the BS, the distance $L_{n}$ varies stochastically
between $L_1$ or $L_2$.}
\end{center} \end{figure}

For paraxial trajectories, where the angle of propagation relative
to the axis is taken to be very small (i.e. $\sin(\theta)\cong
\tan(\theta) \cong \theta$), the reference periodic trajectory
coincides with the optical axis and the monodromy matrix is
identical to the ABCD matrix of the system. The ABCD matrix or
paraxial map of an optical system is the simplest model one can
use to describe the discrete time evolution of a ray in the
optical system \cite{SiegmanBook}. Perhaps the most interesting
and important application of ray matrices comes in the analysis of
periodic focusing (PF) systems in which the same sequence of
elements is periodically repeated many times down in cascade. An
optical cavity provides a simple way of recreating a PF system,
since we can think of a cavity as a periodic series of  lenses
(see Fig \ref{fig:2}). In the framework of geometric ray optics,
PF systems  are classified, as are optical cavities, as  either
stable or unstable.

Without essential loss of generality we restrict ourselves to the
case of a symmetric cavity (i.e. two identical spherical mirrors
of radius of curvature $R$). We take the SOS coincident with the
surface of the BS. After intersecting a given reference plane
$z_{i}$, a ray is transmitted (reflected), it will
 undergo a free propagation over a distance $L_{2}$ ($L_1$), followed by a reflection on the curved mirror $M_2$ ($M_1$), and
 continue  propagating  over the distance $L_2$ ($L_1$), to hit the surface of the beam splitter again at $z_{i+1}$.
 In Fig \ref{fig:2} the sequence of $z_{i}$ represents the successive reference planes  after a
round trip. In the paraxial approximation
 each round trip (time evolution between two successive intersections of a ray
 with the beam splitter) is represented by:

\begin{equation}
\label{eq:1}
\begin{array}{cc}
q_{n+1} =& A_{n}q_{n} +B_{n}\theta_{n}, \\
\theta_{n+1}=& Cq_{n} +D_{n}\theta_{n}, \\
\end{array}
\end{equation}
where
\[ A_{n}=1-2L_{n}/R,\hspace{10mm} B_{n}=2L_{n}(1-L_{n}/R),\]
\[C=-2/R, \hspace{10mm} D_{n}=1-2L_{n}/R \] and
\[L_{n}=\frac{L+p_{n}a}{2}.\]
We have defined $L=L_{1}+L_{2}$ and $a=L_{2}-L_{1}$;  the
stochastic parameter $p_{n}=\pm1$ determines whether the ray is
transmitted ($p_{n}=1$)
or is reflected ($p_{n}=-1$).\\

The  elements of the ABCD matrix  depend on $n$ because of the
stochastic response of the BS, which determines the propagation
for the ray in subcavities of different length (either $L_1$ or
$L_2$). In this way a random sequence of reflections ($p_{n}=1$)
and transmissions ($p_{n}=-1$) represents a particular geometrical
realization of a focusing system. If we want to study the
evolution of a set of rays injected in the cavity with different
initial conditions $(q_0,\theta_0)$, we have two possibilities,
either use the {\em same} random sequence of reflections and
transmissions for all rays in the set or use a {\em different}
random sequence for each ray. In the latter case, we are basically
doing an ensemble average over different geometrical
configurations of focusing systems. As we shall see later it is
convenient, for computational reasons, to adopt the second method.

In the next section we report several dynamical quantities that we
have numerically calculated for paraxial rays in  this system,
using the map described above (Eq.\ref{eq:1}) . The behavior of
these quantities, namely, the SOSs, the exit basins, the Lyapunov
exponent and the escape rate, is analyzed  as a function of the
displacement $(\Delta)$ of the BS with respect to the center of
the cavity (see Fig.\ref{fig:1}).

\section{Results}

The paraxial map of Eq.\ref{eq:1} describes an unbounded system,
that is rays are allowed to go infinitely far from the cavity
axis. In order to describe a physical paraxial cavity we have to
keep the phase space bounded, i.e. it is necessary to artificially
introduce boundaries for the position and the angle of the ray
\cite{Schneider02a}. The phase space boundaries that we have
adopted to decide whether a ray has escaped after a number of
bounces or not is the beam waist ($w_{0}$) and the diffraction
half-angle ($\Theta_0$) of a gaussian beam confined in a globally
stable two-mirror cavity. Measured at the center of the cavity,
$w^{2}_{0}=\frac{L\lambda_{Light}}{\pi}\sqrt{\frac{2R-L}{4L}}$ and
the corresponding diffraction half-angle
$\Theta_{0}=\arctan(\frac{\lambda_{Light}}{\pi w_0})
$\cite{SiegmanBook}. For our cavity configuration we assume
$R=0.15$m , $L=0.2$m and $\lambda_{Light}=500$nm, from which
follows that $w_{0}=5.3 \times10^{-5}$m and
$\Theta_{0}=0.15\times10^{-3}$rad. One should keep in mind that
this choice is somewhat arbitrary and other choices are certainly
possible. The effect of this arbitrariness
on our results will be discussed in detail in section D.\\

\subsection{Poincar\'{e} surface of section  (SOS)}

\begin{figure} \begin{center}
\includegraphics[angle=0,width=17.5truecm]{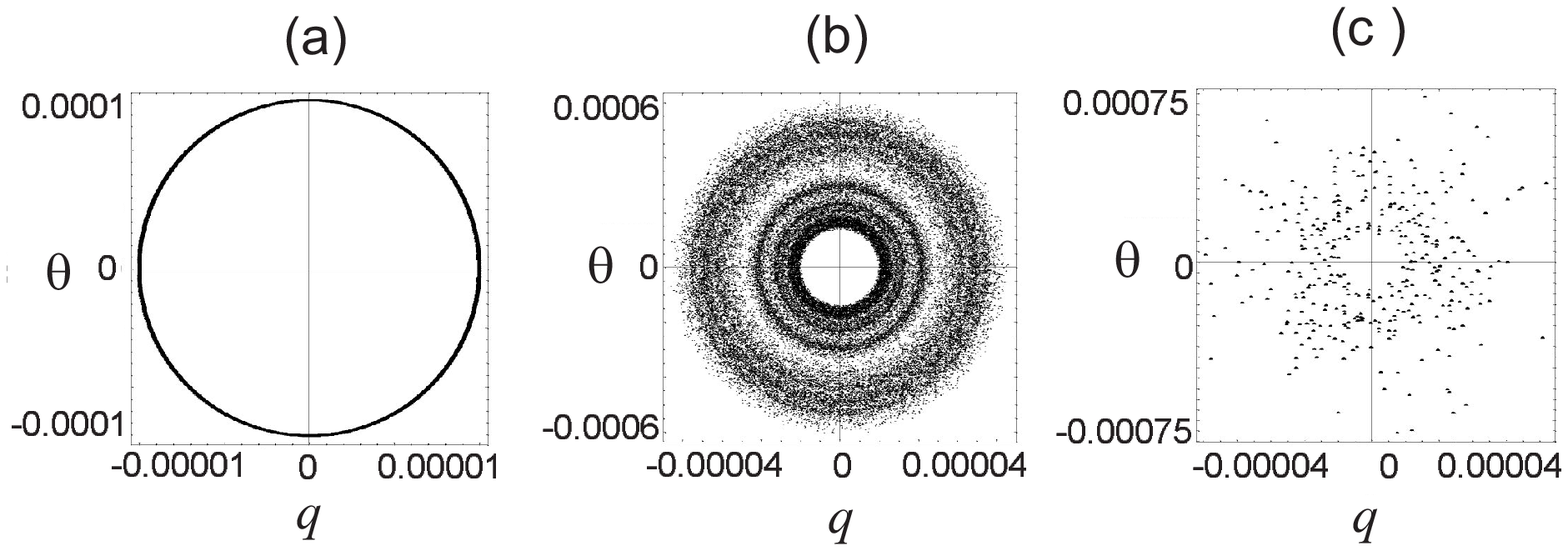}
\caption{\label{fig:4} SOS for (a) $\Delta=0$ the ray does not
escape, (b) $\Delta=0.001$, the ray escapes after $n=5\times
10^{4}$ bounces and (c) $\Delta=0.02$, the ray escapes after
$n=165$.}
\end{center} \end{figure}

We have first calculated the SOS for different positions of the
BS. In order to get a qualitative idea of the type of motion, we
have chosen as transverse phase space variables $y=q$ and
$v_{y}=\sin(\theta)\approx\theta$. The successive intersections of
a trajectory with initial transverse coordinates
$q_0=1\times10^{-5}$m, $\theta_0=0$ are represented by the
different black points in the surface of section. The different
SOSs are shown in Fig \ref{fig:4}. In Fig.\ref{fig:4} (a) we show
the SOS for $\Delta=0$, while in (b) $\Delta=1\times10^{-3}$m and
in (c) $\Delta=2\times10^{-2} $m. In (a) it is clear that the
motion is completely regular (non-hyperbolic); the on-axis
trajectory represents an elliptic fixed point for the map. In (b),
where the BS is slightly displaced from the center
($\Delta=1\times10^{-3}$m) we can see that this same trajectory
becomes unstable because of the presence of the BS, and spreads
over a finite region of the phase space to escape after a large
number of bounces ($n=5\times10^{4}$). In this case we may qualify
the motion as azimuthally ergodic. The fact that the ray-splitting
mechanism introduced by the BS produces ergodicity is a well known
result \cite{Couchman92a} for a closed billard. We find here an
analogue phenomenon, with the difference that in our case the
trajectory does not explore uniformly (but only azimuthally) the
available phase space, because the system is open. Finally, in (c)
we see that the fixed point in the origin becomes hyperbolic, and
the initial
orbit escapes after relatively few bounces ($n=165$). \\

\subsection{Exit basin diagrams}

\begin{figure} \begin{center}
\includegraphics[angle=0,width=17.5truecm]{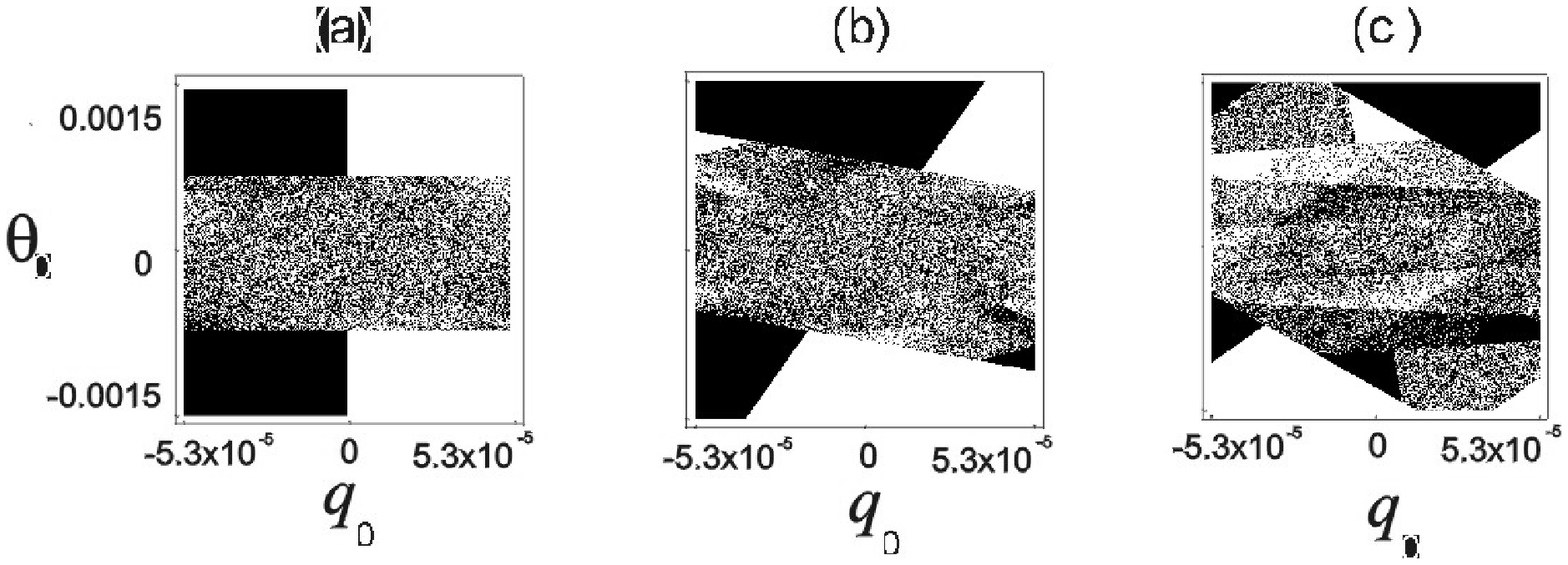}
\caption{\label{fig:5} Exit basins for (a)$\Delta=0.025$,
(b)$\Delta=0.05$  and (c) $\Delta=0.075$.}
\end{center} \end{figure}

It is well known that chaotic hamiltonian systems with more than
one exit channel exhibit irregular escape dynamics which can be
displayed, e.g., by plotting the exit basin \cite{Bleher88a}. For
our open system we have calculated the exit basin diagrams for
three different positions of the BS (Fig.\ref{fig:5}). These
diagrams can be constructed by defining a fine grid
($2200\times2200$) of initial conditions $(q_0,\theta_0)$. We then
follow each ray for a sufficient number of bounces so that it
escapes from the cavity. When it escapes from above
($\theta_{n}>0$) we plot a black dot in the corresponding initial
condition, whereas when it escapes from
below ($\theta_{n}<0$) we plot a white dot.\\
In Fig.\ref{fig:5} (a) we show the exit basins for
$\Delta=0.025$m, the uniformly black or white regions of the plot
correspond to rays which display a regular dynamics before
escaping, and the dusty region  represents the portion of phase
space where there is sensitivity to initial conditions. In Fig.
\ref{fig:5} (b), we
show the same plot for $\Delta=0.05$m, and in (c) for $\Delta=0.075$m.\\
The exit basins plots in Fig.\ref{fig:5} illustrate how the
scattering becomes more irregular as the BS is displaced from the
center. In particular, we see how regions of regular and irregular
dynamics become more and more interwoven as $\Delta$ increases.
Instead, for small values of $\Delta$ as in Fig \ref{fig:5}(a), we
can see that there is a single dusty region  with a uniform
distribution of white and black dots in which no islands of
regularity are present.

\subsection{Escape rate  and Lyapunov exponent}

 The next dynamical quantities we have calculated are the
escape rate $\gamma$ and the Lyapunov exponent $\lambda$. The
escape rate  is a quantity that can be used to measure the degree
of openness of a system \cite{Schneider02a}. For hard chaotic
systems (hyperbolic), the number  $N_{n}$ of orbits still
contained in the phase space  after a long time (measured in
number of bounces $n$) decreases as $N_{0}\exp(-\gamma n)$, while
for soft chaotic systems, the stickiness to
Kolmogorov-Arnold-Moser (KAM) islands (or islands of stability)
leads to  a power law  decay $N_{0}n^{-\gamma}$ \cite{Aguirre03a}.
The Lyapunov exponent is the rate of exponential divergence
of nearby trajectories. \\
Since both $\lambda$ and $\gamma$ are asymptotic quantities they
should be calculated for very long times. In our system long
living trajectories are rare, and in order to pick them among the
grid of initial conditions $N_0$  one has to increase $N_0$ beyond
the computational capability. To overcome this difficulty we
choose a different random sequence for each initial condition. In
this way we greatly increase the probability of picking long
living orbits given by particularly stable random sequences. These
long living orbits in turn make possible the calculation of
asymptotic quantities such as $\lambda$ or $\gamma$.\\

\begin{figure} \begin{center}
\includegraphics[width=9.5truecm]{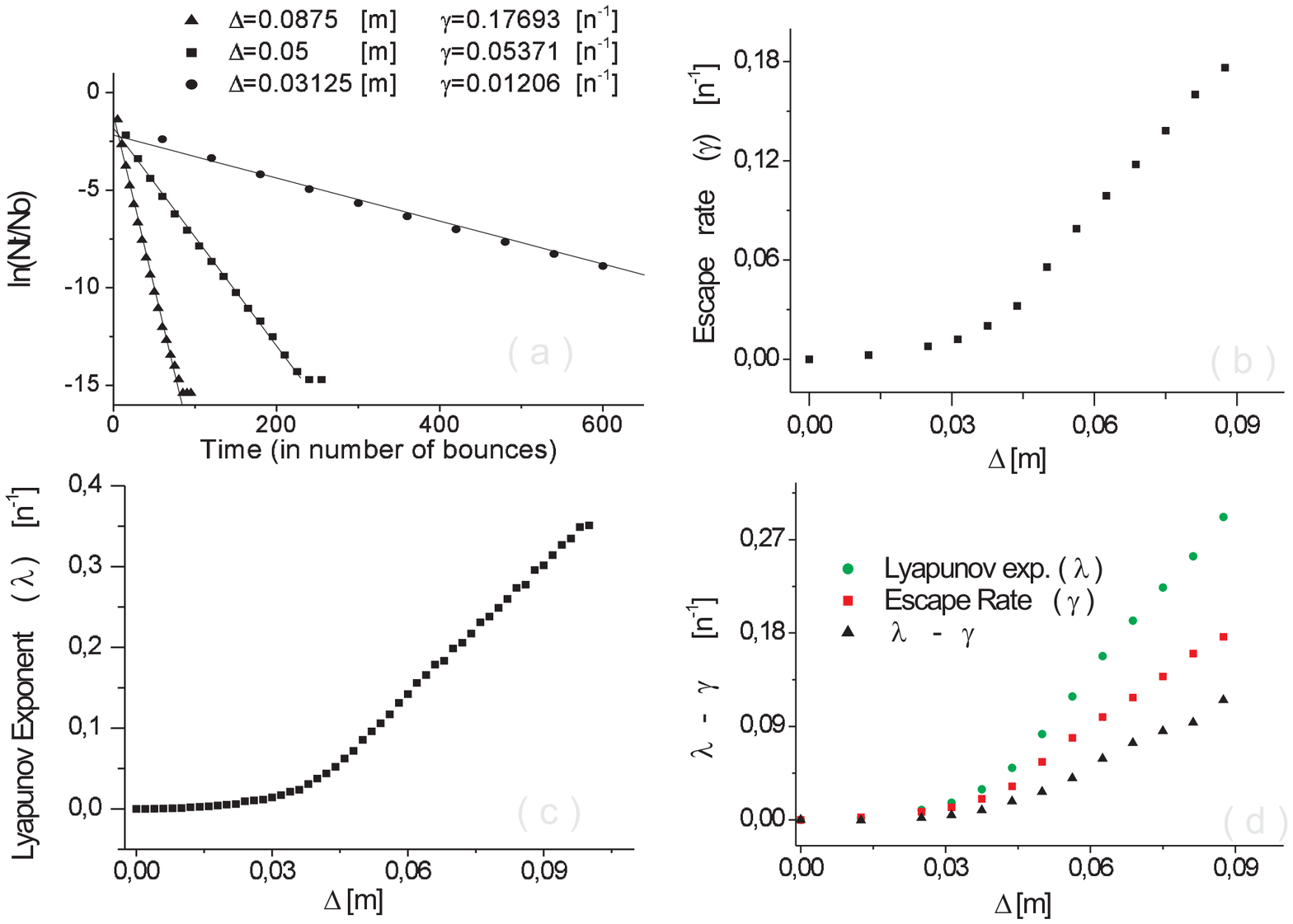}
\caption{\label{fig:6} (a) Linear fits used to calculate the
escape rate for three different geometrical configurations of the
cavity given by $\Delta=0.03125$m, $\Delta=0.05$m and
$\Delta=0.0875$m. The time is measured in number of bounces [$n$].
The slope $\gamma$ is in units of the inverse of time [$n^{-1}$].
Fig (b) shows the escape rate $\gamma$ [$n^{-1}$] as a function of
$\Delta$. Fig. (c) corresponds to different Lyapunov exponents
$\lambda$ [$n^{-1}$] as the BS moves from the center $\Delta=0$ to
the leftmost side of the cavity $\Delta=0.10$m. Fig. (d) shows the
difference between $\lambda-\gamma$ [$n^{-1}$], which is a
positive bounded function. }
\end{center} \end{figure}

The escape rate $\gamma$ was determined measuring $N_{n}$, as the
slope of a linear fit in the $N_{n}/N_{0}$ versus $n$ curve, in a
logarithmic scale; the total number of initial conditions $N_0$
being chosen as $2200\times2200$.\\
We have calculated the dependence of $\gamma$  with the
displacement $\Delta$ of the BS from the center of the cavity,
where $0\leq\Delta\leq L/2$. Since for $\Delta > R - L/2$ the left
subcavity becomes unstable, it would seem natural to expect that
this position of the BS would correspond to a critical point.
However, we have found by explicit calculation of both the
Lyapunov exponent and the escape rate, that such a critical point
does not manifest itself in a sharp way, rather we have observed a
finite transition region (as opposed to a single point) in which
the functional dependence of $\lambda$ and $\gamma$ change in a
smooth way. In Fig \ref{fig:6} (a) we show the typical behavior of
$\frac{N_{n}}{N_{0}}$vs $n$ in semi-logarithmic plot for three
different positions of the BS. The displacement of the BS is
$\Delta = 0.0875$m,  $0.05$m and $0.03125$m, and the corresponding
slopes (escape rate $\gamma$ measured in units of the inverse
number of bounces $n$) of the linear fit are $\gamma =
0.17693n^{-1}$, $0.05371n^{-1}$ and $0.01206n^{-1}$ respectively.
We have found that the decay is exponential only up to a certain
time (approximately 70-1000 bounces depending on the geometry of
the cavity) due the
discrete nature of the grid of initial conditions.\\
In Fig \ref{fig:6} (b) we see that $\gamma$ increases with
$\Delta$, revealing that for more unstable configurations there is
a higher escape rate, as  expected.  Its also interesting to
notice that the exponential decay fits better when the beam
splitter is further from the center position, since this leads to
smaller stability of the periodic orbits of the system. However,
the dependence of the escape rate with the position of the BS is
smooth and reveals that the  only critical displacement, where the
escape rate becomes positive, is  $\Delta=0$.

As a next step, we have calculated the Lyapunov exponent $\lambda$
for the paraxial map; $\lambda$ is a quantity that measures the
degree of stability of the reference periodic orbit. For a
two-dimensional  hamiltonian map there are two  Lyapunov exponents
($\lambda_1$, $\lambda_2$) such that $\lambda_1 + \lambda_2=0$. In
the rest of the paper we shall indicate with $\lambda$ the
positive Lyapunov exponent which quantifies the  exponential
sensitivity to the initial conditions. We have calculated
$\lambda$ for the periodic orbit on axis, using the standard
techniques \cite{Benettin78a}, and we have found that the Lyapunov
exponent grows from zero  with the distance of the BS to the
center (Fig \ref{fig:6} (c)). Therefore, the only critical point
revealed by the ray dynamics is again the center of the cavity
($\Delta=0$), where the magnitudes change from zero to a positive
value. This result also shows that  the presence of the BS with
its stochastic nature introduces exponential sensitivity to
initial conditions in the system for every $\Delta\neq0$, even
when both subcavities are stable. This surprising fact can be
explained by taking into account the well known probabilistic
theorem by Furstenberg on the asymptotic limit of the rate of
growth of a product of random matrices (PRM) \cite{Furstenberg}.
From this theorem we  expect that the asymptotic behavior of the
product $M_{n}$ of a uniform sequence $\omega$ of independent,
random, unimodular, $D \times D$ matrices, and for any nonzero
vector $\vec{y}\in \Re^D$:
\begin{equation}
\label{eq:2}\lim_{n\rightarrow\infty} \frac{1}{n} \langle \ln
|M_{n} \vec{y}| \rangle = \lambda_{1} > 0,
\end{equation}\\
where $\lambda_1$ is the maximum  Lyapunov characteristic exponent
of the system, and the angular bracket indicates the average over
the ensemble $\Omega$ of all possible  sequences $\omega$. This
means that for PRM the Lyapunov exponent is a nonrandom positive
quantity. In general, it can be said that there is a subspace
$\Omega^*$ of random sequences which has a full measure
(probability 1) over the whole space of sequences $\Omega$ for
which nearby trajectories deviate exponentially at a rate
$\lambda_1$. Although there exist very improbable sequences in
$\Omega$ which lead to a different asymptotic limit, they do not
change the logarithmic average (Eq.\ref{eq:2})\cite{PRM}. We have
verified this result, calculating the value of $\lambda$ for
different random sequences $\omega_{i}$, in the asymptotic limit
$n=100000$ bounces, and we obtained in all cases the same Lyapunov
exponent.
\subsection{Mixing properties}
Dynamical randomness  is characterized by a positive
Kolmogorov-Sinai (KS) entropy per unit time $h_{KS}$
\cite{Gaspard90a}. In closed systems, it is known that dynamical
randomness is a direct consequence of the exponential sensitivity
to initial conditions given by a positive Lyapunov exponent. On
the other hand, in open dynamical systems with a single Lyapunov
exponent $\lambda$, the exponential sensitivity to initial
conditions can be related to $h_{KS}$ through the escape rate
$\gamma$, by the relation \cite{GaspardBook}:
\begin{equation}\label{eq:100}
  \lambda = h_{KS}+ \gamma.
\end{equation}
This formula reveals the fact that in an open dynamical system the
exponential sensitivity to initial conditions induces two effects:
one is the escape of trajectories out of the neighborhood of the
unstable reference periodic orbit at an exponential rate $\gamma$,
and the other one is a dynamical randomness  because of transient
chaotic motion near this unstable orbit \cite{GaspardBook}. This
dynamical randomness is a measure of the degree of mixing of the
system and as mentioned before is quantified by $h_{KS}$.
Therefore, for a given $\lambda$, the larger the mixing is, the
smaller the escape rate, and vice versa. From
Figs.\ref{fig:6}(b,c) it is evident that the Lyapunov exponent and
the escape rate have the same smooth dependence on the BS
displacement $\Delta$ and that $\gamma\leq\lambda$. We have
calculated the difference $\lambda-\gamma>0$ for our system and
the result is shown in Fig\ref{fig:6} (d).\\
The actual value of $\gamma(\Delta)$ depends, for a fixed value of
$\Delta$, on the size of the phase space accessible to the system
\cite{Schneider02a}, that is, it depends on $w_0$ and $\theta_0$.
We verified this behavior by successively decreasing $w_0$ and
$\theta_0$ by factors of 10 (see Table \ref{tab:1}), and
calculating $\gamma$ for each of  these phase space boundaries. It
is clear from these results that $\gamma$ increases when the size
of phase space decreases; in fact for $w_0, \theta_0 \approx 0$,
one should get $\lambda\approx \gamma$ and the cavity mixing
property should disappear. It is important to notice that the
increment of $\gamma$ with the inverse of the size of the
accessible phase space is a general tendency, independent from the
arbitrarily chosen boundaries.

\begin{table}[h!]
\begin{center}
\begin{tabular}{|c|c|c|c|c|}
\hline
  ($w_0,\theta_0$) & $\times10^{0}$ & $\times10^{-1}$ & $\times10^{-2}$ & $\times10^{-3}$ \\
\hline \hline
  $\gamma$ &  0.17639 & 0.17596 & 0.19559  & 0.25259 \\ \hline
\end{tabular}
\caption{\label{tab:1} Escape rate for different phase space
boundaries. As the boundary shrinks $\gamma(\Delta)$ tends to the
corresponding value of $\lambda(\Delta)=0.29178n^{-1}$. In these
calculations the displacement of the BS was $\Delta=0.0875$m.}
\end{center}
\end{table}

It is important to stress that, although the randomness introduced
by the stochastic BS is obviously independent from the cavity
characteristics, $\lambda$ and $\gamma$ show a clear dependance on
the BS position. When the BS is located at the center of the
cavity it is evident for geometrical reasons that the ray
splitting mechanism becomes ineffective: $\lambda = 0 = \gamma$.
These results confirm what we have already shown in the SOS (Fig.
4).\\

\section{CONCLUSIONS}

We have been able to characterize the ray dynamics of our optical
cavity with ray splitting by using standard techniques in
nonlinear dynamics. In particular we have found, both through the
SOS and the exit basin diagrams, that the stochastic ray splitting
mechanism destroys the regular motion of rays in the globally
stable cavity. The irregular dynamics introduced by the beam
splitter was quantified by calculating the Lyapunov exponent
$\lambda$; it grows from zero as the beam splitter is displaced
from the center of the cavity. Therefore, the center of the cavity
constitutes the only point where the dynamics of the rays is not
affected by the stochasticity of the BS. The escape rate $\gamma$
has been calculated and it has revealed a similar dependence with
the position of the beam splitter to that of $\lambda$.
Furthermore, we have verified that the absolute value of the
escape rate tends to that of the Lyapunov exponent as the size of
the available phase space goes to zero. This result confirms the
fact that the escape rate and therefore the mixing properties of a
map depend sensitively on the choice of the boundary
\cite{Schneider02a}. Because of this dependence we cannot claim
that our system is chaotic, despite the positiveness of $\lambda$.
However, in a future publication we shall demonstrate that ray
chaos \textit{can} be achieved for the same class of optical
cavities when \textit{non}-paraxial ray dynamics is allowed
\cite{Puentes}.\\

This project is part of the program of FOM and is also supported
by the EU under the IST-ATESIT contract. We thank S. Oemrawsingh
for useful contributions to software development.

\newpage
\newpage
\newpage
\newpage
\newpage
\newpage
\newpage
\newpage


\begin{thebibliography}{ab16}

\bibitem{Kohler97a}A. Kohler, G. H. Killesreiter, and R.
Bl\"{u}mel, Phys. Rev. E \textbf{56}, 2691 (1997).

\bibitem{Couchman92a}L. Couchman, E. Ott, and T. M. Antonsen, Jr.,
Phys. Rev. A \textbf{46}, 6193 (1992).

\bibitem{Hentschel02a}M. Hentschel and K. Richter, Phys.\ Rev. E
\textbf{66}, 056207 (2002).

\bibitem{SiegmanBook}A. E. Siegman, \textit{Lasers} (University
Science Books, Mill Valley, CA, 1996).

\bibitem{Cvitanovic02}P. Cvitanovi\'{c} \textit{et al., Classical
and Quantum Chaos} (www.nbi.bk/ChaosBook/, 2002).

\bibitem{Longhi}S. Longhi, Phys. Rev. E \textbf{65}, 027601
(2002).

\bibitem{OttBook}E. Ott. \textit{Chaos in Dynamical Systems}
(Cambridge University Press, 2002), 2nd ed.

\bibitem{Schneider02a}J. Schneider, T. T\'{e}l, and Z. Neufeld,
Phys. Rev. E \textbf{66}, 066218 (2002).

\bibitem{Bleher88a}S. Bleher, C. Grebogi, E. Ott, and R. Brown,
Phys. Rev. A \textbf{38}, 930 (1988).

\bibitem{Aguirre03a}J. Aguirre and M. A. F. Sanju\'{a}n, Phys.
Rev. E \textbf{67}, 056201 (2003).

\bibitem{Benettin78a}G. Benettin and J. M. Strelcyn, Phys. Rev. A
\textbf{17}, 773 (1978).

\bibitem{Furstenberg}H. Fustenberg, Trans. Amer. Math. Soc.
\textbf{108}, 377 (1963).

\bibitem{PRM}A. Crisanti, G. Paladin and A. Vulpani, \textit{Products of Random
Matrices} (Springer-Verlag, 1993).

\bibitem{Gaspard90a}P. Gaspard and G. Nicolis, Phys. Rev. Lett.
\textbf{65}, 1693 (1990).

\bibitem{GaspardBook}P. Gaspard, \textit{Chaos, Scattering and Statistical
Mechanics}(Cambridge University Press, 1998), 1st ed.


\bibitem{Puentes}G. Puentes, A. Aiello, and J. P. Woerdman, \textit{in
preparation} (2003).













\end{thebibliography}
\end{document}